\documentclass[useAMS,usenatbib]{mn2e}
\usepackage{graphicx}
\newcount\longrefs
\def\aap{\ifnum\longrefs=1 {Astron.\ Astrophys.}\else
                           {A\hbox{\rm \&}A}\fi}
\def\aapr{\ifnum\longrefs=1 {Astron.\ Astrophys.\ Rev.}\else
                            {A\hbox{\rm \&}AR}\fi}
\def\aaps{\ifnum\longrefs=1 {Astron.\ Astrophys.\ Suppl.}\else
                            {A\hbox{\rm \&}A Suppl.}\fi}
\def\aj{\ifnum\longrefs=1 {Astron.\ J.}\else
                          {AJ}\fi}
\def\ao{\ifnum\longrefs=1 {Applied Optics}\else
                           {Appl.\ Opt.}\fi}
\def\aspcs{\ifnum\longrefs=1 {Astron.\ Soc.\ Pacific Conf. Series}\else
                           {ASP Conf.\ Ser.}\fi}
\def\apj{\ifnum\longrefs=1 {Astrophys.\ J.}\else
                           {ApJ}\fi}
\def\apjl{\ifnum\longrefs=1 {Astrophys.\ J. Lett.}\else
                            {ApJ}\fi}
\def\aplett{\ifnum\longrefs=1 {Astrophys.\ J. Lett.}\else
                            {ApJ}\fi}
\def\apjs{\ifnum\longrefs=1 {Astrophys.\ J. Suppl.}\else
                            {ApJS}\fi}
\def\apss{\ifnum\longrefs=1 {Astrophys.\ and Space Science}\else
                            {Astrophys.\ Space Sci.}\fi}
\def\araa{\ifnum\longrefs=1 {Ann.\ Rev.\ Astron.\ Astrophys.}\else
                            {ARA\hbox{\rm \&}A}\fi}
\def\azh{\ifnum\longrefs=1 {Astronomicheskii Zhurnal}\else
                            {Astron.\ Zhur.}\fi}
\def\baas{\ifnum\longrefs=1 {Bull.\ Am.\ Astron.\ Soc.}\else
                            {BAAS}\fi}
\def\bain{\ifnum\longrefs=1 {Bull.\ Astronom.\ Institutes Netherlands}\else
                            {Bull.\ Astr.\ Inst.\ Neth.}\fi}
\def\gca{\ifnum\longrefs=1 {Geochim.\ Cosmochim.\ Acta}\else
                           {Geochim.\ Cosmochim.\ Acta}\fi}
\def\grl{\ifnum\longrefs=1 {Geophys.\ Res.\ Lett.}\else
                           {Geoph.\ Res.\ Lett.}\fi}
\def\iaucirc{\ifnum\longrefs=1 {IAU Circulars}\else
                          {IAU Circ.}\fi}
\def\ip{\ifnum\longrefs=1 {in press}\else
                          {in press}\fi}
\def\jgr{\ifnum\longrefs=1 {J.\ Geophys.\ Res.}\else
                           {J.\ Geophys.\ Res.}\fi}
\def\jrasc{\ifnum\longrefs=1 {J.\ Royal Astron.\ Soc.\ Canada}\else
                           {JRAS Can.}\fi}
\def\memsai{\ifnum\longrefs=1 {Mem.~Soc.~Astron.~Italiana}\else
                              {MemSAI}\fi}
\def\mnras{\ifnum\longrefs=1 {Mon.\ Not.\ Roy.\ Astron.\ Soc.}\else
                             {MNRAS}\fi}
\def\nat{\ifnum\longrefs=1 {Nature}\else
                           {Nat}\fi}
\def\pasj{\ifnum\longrefs=1 {Pub.\ Astron.\ Soc.\ Japan}\else
                            {PASJ}\fi}
\def\pasp{\ifnum\longrefs=1 {Pub.\ Astron.\ Soc.\ Pacific}\else
                            {PASP}\fi}
\def\physscr{\ifnum\longrefs=1 {Physica Scripta}\else
                            {Phys.\ Scrip.}\fi}
\def\planss{\ifnum\longrefs=1 {Planetary \& Space Science}\else
                            {Plan. \& Space Sci.}\fi}
\def\procspie{\ifnum\longrefs=1 {Proc.\ SPIE}\else
                            {Proc.\ SPIE}\fi}
\def\qjras{\ifnum\longrefs=1 {Quarterly J.\ Royal Astron.\ Soc.}\else
                            {QJRAS}\fi}
\def\sa{\ifnum\longrefs=1 {Soviet Astron..}\else
                               {Sov.\ Astron.}\fi}
\def\skytel{\ifnum\longrefs=1 {Sky \& Telescope}\else
                            {Sky \& Tel.}\fi}
\def\solphys{\ifnum\longrefs=1 {Solar Phys.}\else
                               {Sol.\ Phys.}\fi}
\def\ssr{\ifnum\longrefs=1 {Space Science Rev.}\else
                               {Space\ Sci.\ Rev.}\fi}
\def\zap{\ifnum\longrefs=1 {Zeitschr.\ f.\ Astrophysik}\else
                               {Z.\ Astrophys.}\fi}
 \def\kms{km s$^{-1}$}
 \def\cm1{$\rm cm^{-1}$}
\def\kms{$\rm km\,s^{-1}$}
\def\DE{D\kern-0.75em \raisebox{1.0pt}{=}\ }

\def\I{{\sc i}}

\title[Abundance sensitive points of line profiles]
{Abundance sensitive points of line profiles\\ in the stellar spectra}
\author[V. A. Sheminova and C. R. Cowley]
       {V. A. Sheminova$^{1}$\thanks{E-mail:shem@mao.kiev.ua (VAS)}
    and C. R. Cowley$^{2}$\thanks{E-mail:cowley@umich.edu (CRC)}\\
$^{1}$Main Astronomical Observatory,
                National Academy of Sciences of the Ukraine,
                27 Akademika Zabolotnoho St.,
                03680 Kiev, Ukraine\\
$^{2}$Department of Astronomy,
                 University of Michigan,
                 Ann Arbor, MI 48109-1042, USA  }
\date{Accepted 2013 Month  00. Received 2013 Month 00; in original form 2013 Month 00}

\begin{document}

\maketitle

\begin{abstract}
Many abundance studies are based on spectrum synthesis and $\chi$-squared
differences between the synthesized and an observed spectrum.  Much of the
spectra so compared depend only weakly on the elemental abundances.
Logarithmic plots of line depths rather than relative flux make this more
apparent. We present simulations that illustrate a simple method for finding
regions of the spectrum most sensitive to abundance, and also some caveats
for using such information. As expected, we find that weak features are the
most sensitive. Equivalent widths of weak lines are ideal features, because
of their sensitivity to abundances, and insensitivity to factors that
broaden the line profiles.  The wings of strong lines can also be useful,
but it is essential that the broadening mechanisms be accurately known.  The
very weakest features, though sensitive to abundance, should be avoided or
used with great caution because of uncertainty of continuum placement as
well as numerical uncertainties associated with the subtraction of similar
numbers.

\end{abstract}


\section{Introduction}

Since the early days of analytical stellar spectroscopy, it has been known
that equivalent widths have numerous advantages over the use of profiles. It
is easily demonstrated that equivalent widths are unaffected by various and
often uncertain broadening mechanisms, both stellar and instrumental (see
below). Nevertheless, modern methods have made it possible to synthesize
large regions of the spectra of many stars. Following the pioneering paper
by \citet{1996ASPC..108..175V}, 
{\em Spectroscopy Made Easy}, considerable work has been based on automated
methods,  some based on equivalent widths, but many stressing spectral
synthesis (cf.
 \citealt{2014A&A...564A.109S}, 
 \citealt{Blanco}, 
and references therein). While these methods have been demonstrated to work
quite well, they could be strengthened {\em by incorporating knowledge of
wavelengths sensitive to the stellar properties sought}.

We emphasize abundances here.  In practice other factors ($T_e$, $\log(g)$,
microturbulence) must be known, and generally must be sought simultaneously
with abundances.  This is a strength of some of the automated methods.  The
basic technique we advocate could be adapted for these other parameters, but
that is not done here.

For the most part, automated methods are used with cooler stars, where the
abundance patterns are of limited scope, e.g.  [Fe/H] or [$\alpha$/Fe]
variations. The situation is more subtle with chemically peculiar stars (CP)
of the upper main sequence, or peculiar red giants, where element-to-element
variations can be large. The analysis of young stars are also problematical,
where standard, LTE models are of questionable validity, especially in the
region where the cores of strong lines are formed.

Our goal is to show that knowledge of abundance-sensitive regions of line
profiles make it possible to improve accuracy.  We illustrate how
inaccuracies can arise from the use of regions of line profiles less
sensitive to abundance than other factors such as turbulence, or damping.

\section[]{Sensitivity of the profile points to abundance}

It is convenient to consider changes in line profiles resulting from
abundance variations one element at a time. In the present examples the
element will be iron.  We use the symbol $A =12+\log \rm(Fe/N_{tot})$ for
abundances on the usual logarithmic scale, where for hydrogen $A = 12.00$,
 $\rm Fe/N_{tot}$ is the number of iron atoms and ions to the sum for all
elements including iron.  The illustrations are all simulations where the
input parameters are precisely known.  We vary abundance, microturbulence
($\xi_t$), rotation ($V\cdot\sin(i)$), and the macroturbulence ($V_{\rm
mac}$). The latter is taken to be a Gaussian, and assumed here to include
instrumental broadening.  In most stellar work, the macroturbulence is
assumed to be isotropic with a Gaussian profile. Gaussians are also
typically assumed for instrumental profiles.

Let the parameter   defining   the  profile of an absorption line be the
line depth,
 $R_{\lambda}=(F_c-F_{\lambda})/F_c$. Here $F_{\lambda}$ and $F_c$ are
the flux at the wavelength $\lambda$ in the line profile and continuum,
respectively. Following  \citet{1995ASPC...81..467C}  the sensitivity of
each profile point to the $A$-variations may be calculated by the ratio

 \[
S_{\lambda} =100\cdot \frac{ R_{\lambda}(A+\Delta A) - R_{\lambda}(A-\Delta
A)}{R_{\lambda}(A)}.
 \]

The spectra are calculated with an assumed abundance $A$ and again with $A$
varied by $\pm\Delta A$. We adopted  $\Delta A= 0.1$ dex to avoid large
changes in the line profiles. Essentially the same information is obtained
with other small values of $\Delta A$.  In this way we can calculate the
sensitivity  of each point of line profile to abundance without involving
the more complex response functions which involve partial derivatives of the
emergent flux with respect to the free parameters
 \citep[see, e.g.,][] {1975SoPh...43..289B, 
 1977A&A....54..227C, 
 1993KFNT....9...27S}. 

 \begin{figure}
    \centering
    \includegraphics[width=8.5cm]{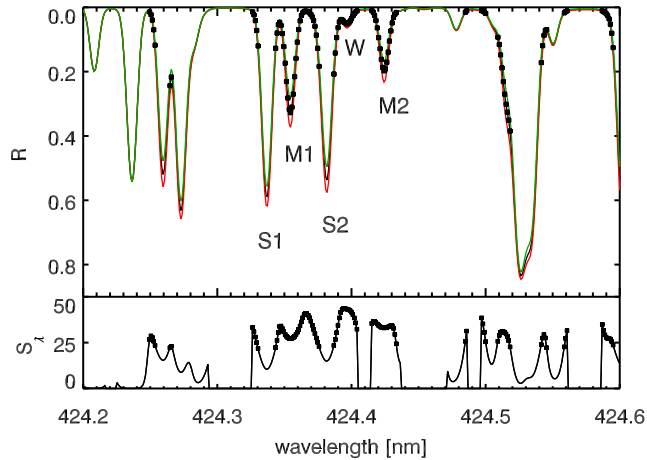}
   \caption{Line depth  $R_{\lambda}=(F_c-F_{\lambda})/F_c$
    calculated with assumed
    iron abundance $A=7.46$ (black),  $A-0.1$ (green),  $A+0.1$ (red).
    The sensitivity function $S_\lambda$ is shown in the subfigure,
    with the most sensitive points indicated.
    Only the iron abundance has been changed.  Here, and in the following
    figures, colors refer to the online version.
      }
  \label {fig:Fig1}
  \end{figure}

Calculations are based on the synthesis code SPANSAT
\citep{1988ITF...87P....3G}, and a MARCS model atmosphere
\citep{2008A&A...486..951G} with $T_{\rm eff} = 5777$~K, $\log(g) = 4.44$,
and the chemical composition of the Sun   \citep{2009ARA&A..47..481A}. Local
thermodynamic equilibrium (LTE) is assumed throughout.  The MARCS model is
available online at http://marcs.astro.uu.se.  We use a depth-independent
microturbulence $\xi_t = 1$~\kms, corresponding to that used in the MARCS
model and an isotropic  macroturbulence  of 2.4~\kms. The rotational
velocity ($V\cdot\sin(i) = 1.85$~\kms \citep{1984ApJ...281..830B}) was
simulated by direct averaging over the disk.  The synthesis of spectral
regions includes the full list of lines available in the VALD database
\citep{1999A&AS..138..119K}, with line parameters: wavelengths, excitation
potentials, oscillator strengths, and damping parameters.

A few features that are in the real solar spectrum are not in our
calculation.  These include molecular lines, not in our version of VALD, as
well as unidentified or unclassified atomic lines.

 \begin{figure}
    \centering
     \includegraphics[width=8.5cm]{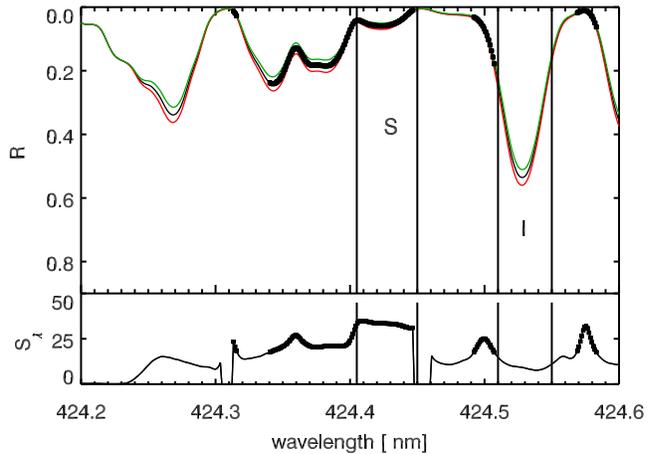}
   \caption{The same as Fig.~\ref{fig:Fig1}  but with relatively
    high rotation velocity $V \sin i =15$~\kms.  Symbols `S' and
    `I' mark regions sensitive and insensitive to the iron
    abundance.
     }
  \label {fig:Fig2}
  \end{figure}
 \begin{figure}
 \centering
  \includegraphics[width=80mm,height=55mm,angle=-0]{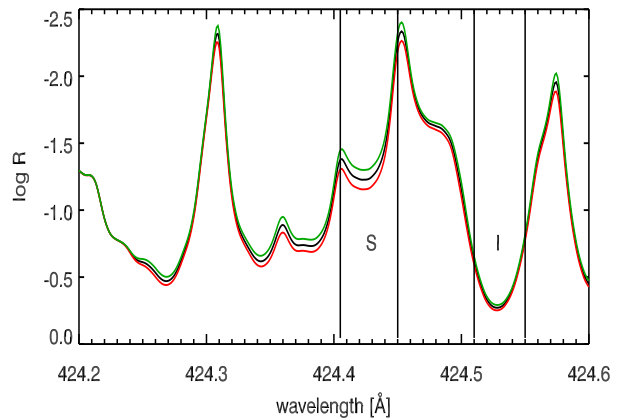}
\caption{This plot was made with different parameters from those of Fig. 2.
The logarithmic display makes the sensitivity of weak features more obvious
than traditional plots of rectified spectra and fitted calculations.
  }
\label {fig:pl3log}
\end{figure}

 \subsection{High resolution and narrow-lined spectrum}

The eye is drawn most naturally to the cores of the stronger lines, where
the separation for the different abundances is most evident. The calculated
sensitivity function shows how deceptive this visual impression can be.  The
cores of the stronger features are generally the {\em least} sensitive parts
of the profile to abundance.

The most sensitive feature in Fig.~\ref{fig:Fig1} is the weak absorption at
424.395 (W), which shows the advantage of using weak lines for abundances.
The two stronger lines, S$_1$ and S$_2$ are sensitive only in their wings.
The moderately strength lines, M$_1$ and M$_2$ show intermediate cases.
While the wings of moderate and strong lines can also be quite sensitive to
abundances, these regions are also sensitive to other, sometimes quite
complicating factors.  Beyond the Doppler core, the wing strength depends on
the product of the abundance, the line strength, and the damping constant.
Hyperfine structure, and Zeeman broadening could be relevant. In addition,
the instrumental profile must be accurately known as well as broadening due
to stellar turbulence (micro and macro), before accurate abundances can be
determined from line wings.

These factors have virtually precluded the practical use of line wings in
most abundance studies.  There is, however, new work in which advantage
could be taken of the sensitivity of line wings.  We refer specifically to
the new differential work on abundances of solar-type stars (e.g.
 \citealt{2009ApJ...704L..66M, 2013arXiv1307.5274M}). 
 Thus far, this work has been based on equivalent widths.
However, in comparing two closely similar stars, with spectra obtained by
identical instruments, and reduced with the same procedures, the unknown
broadening mechanisms should cancel. There is thus good reason to hope that
these differential methods can be strengthened through the use of line
wings.

\subsection{Lower resolution and convoluted profiles}

A great deal of important work deals with convoluted spectra, where it may
not be possible to analyze weak, isolated features.  This convolution may
arise from the use of low instrumental resolution or in studies of
integrated spectra of stellar systems.  Sensitivity functions should be
comparably useful in all of these cases. The spectra of single stars are
often convoluted by rotation, which we discuss here.

The region shown in Fig.~\ref{fig:Fig2} is the same as in ~\ref{fig:Fig1}
but the profiles were broadened by an assumed rotation.  The appearance of
the spectra and the sensitivity functions themselves are markedly changed.
Wider regions of the spectrum are now sensitive, though the sensitivities
themselves are lowered -- in the case of the maxima, by some 21 per cent. It
would surely be useful for an abundance worker to see that the region from
424.41 to 424.45 nm was relatively sensitive (S) to the iron abundance while
that from 424.51 to 424.55 was not (I).

\subsection{Logarithmic plots\label{sec:logs}}

Traditional plots showing observed and calculated spectra obscure the
sensitivity of weak features that is shown by the sensitivity functions.  A
20 \% change of a feature that is only 10 \% deep makes a change in the
depth of only 0.02 in the depth. This difference is hardly noticable on a
standard plot. By contrast, a change of only 5\% on a line of depth 0.9
makes a change of 0.045, more than double the much larger {\em percentage}
change of the weaker feature.

It might be helpful for abundance workers to use plots in which the
logarithms of line depths are plotted rather than linear, relative fluxes.
An example is shown in the  Fig.~\ref{fig:pl3log}, it conveys much the same
information as the lower part of Fig.~\ref{fig:Fig2}. The logarithmic plot
compresses the larger line depths; the (online color) separation in the
sensitive regions are more obvious  while that for the stronger cores are
suppressed. A logarithmic plot similar to Fig.~\ref{fig:pl3log}, but showing
observed and calculated spectra would help to focus attention  on the
weaker, more abundance-sensitive parts of the spectra.  Unlike the
sensitivity functions, such plots do not have the ability to display
sensitivity to specific elements.

  \section{Weak features}
  \label{sec:wkfea}

In this section a number of calculations are presented, based on the Fe I
line at 6151.62~\AA.  Its (lower) excitation potential is 2.18 eV.  The
$\log(gf)$ given by
 \citet{2006nla..conf..278W} 
 is $-$3.29. However, this value is varied for purposes
of illustration. All profiles described in this section were calculated with
the wavelength step of 5~m\AA\ and MARCS solar model.
 \begin{figure}
    \centering
  \includegraphics[width=70mm,height=65mm,angle=-00]{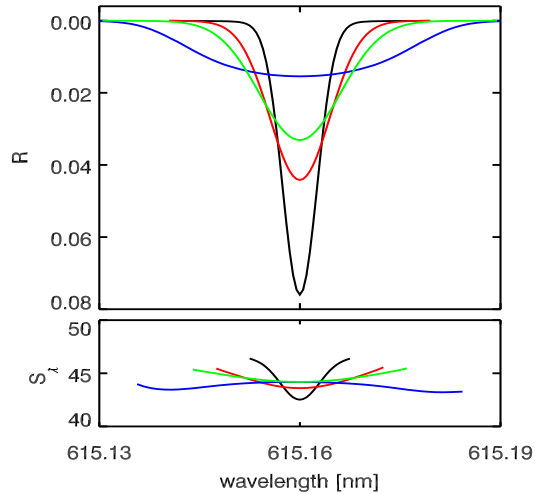}
   \caption{These four profiles all have the same equivalent width,
   and were made with the same abundance.  Only the broadening
   parameters differ (see Tab.~\ref{tab:weaklins}). Lower panel shows
   the sensitive functions of these profiles in their respective colors.
     }
  \label {fig:eqwth}
  \end{figure}

\subsection{Equivalent widths and profiles}

The independence of equivalent widths of weak lines is illustrated in
Fig.~\ref{fig:eqwth}, where very different profiles yield the same
abundances.  This of course is a well known result.   We use it here to
emphasize that in this case, all of the relevant points are sensitive in the
sense used in this paper.  The sensitive function of each profile is high
although it varies slightly with line depth. The parameters used in
Fig.~\ref{fig:eqwth} are given in Table~\ref{tab:weaklins}.
  \begin{table}
   \centering
  \caption{Parameters used for the four profiles of Fig.~\ref{fig:eqwth}.
  The listing is from the narrowest to broadest profile.}
  \label{tab:weaklins}
  \begin{tabular}{c c c c r c} \hline
  Pro. & A    & $\xi_t$ &$V_{\rm mac}$&  $V\cdot\sin(i)$ &W  \\
       & log  & \kms    & m\AA  &  \kms &  m\AA\,  \\
       \hline
    1  & 7.46 &  1.0    &  0.6  &  0.0 &4.96              \\
    2  & 7.46 &  1.4    &  2.4  &  0.0 &5.00              \\
    3  & 7.46 &  2.0    &  3.4  &  0.0 &5.05              \\
    4  & 7.46 &  1.4    &  2.4  & 10.0 &5.01              \\ \hline
 \end{tabular}
 \end{table}

\subsection{Weak-line fits with variable parameters}

 \begin{figure}
    \centering
  \includegraphics[width=85mm,height=85mm,angle=-00]{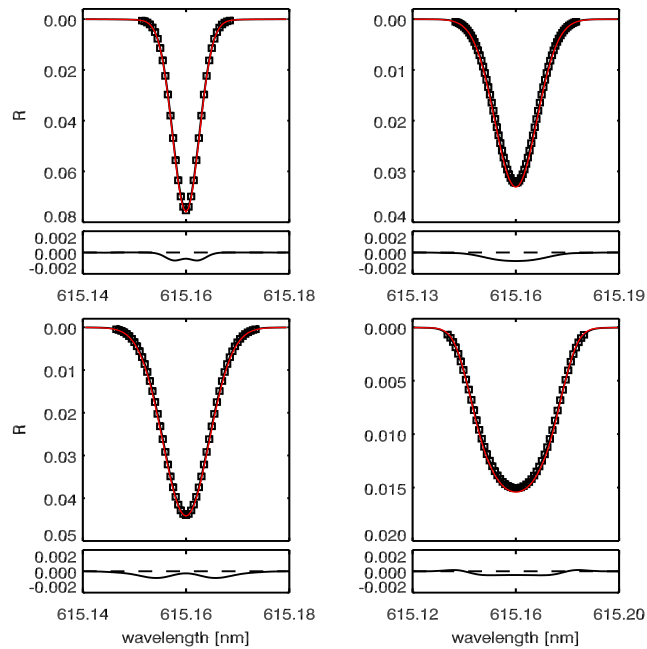}
   \caption{The red profiles are the same as in Fig.~\ref{fig:eqwth}.
   The squares  refer to calculations made with the abundances,
   $\log(gf)$ and $\xi_t$ frozen at 7.46, -4.65, and 0.4,
   respectively, but with $V_{\rm mac}$ adjusted to obtain an
   acceptable fit.  Residuals ($\Delta R=R -R_{\rm original}$) are shown in
   the lower  panels of each subfigure.
     }
  \label {fig:All4}
  \end{figure}

A numerical experiment shows that a close fit to the line profiles does not
necessarily fix the microturbulence uniquely.  Alternately, one may say that
one can get a good abundance from weak lines, even if the microturbulence is
not accurately known.   In Fig.~\ref{fig:All4} we match individually, the
four profiles of Fig.~\ref{fig:eqwth} with altered parameters. The four
profiles of Fig.~\ref{fig:eqwth} are all closely matched with the abundance
(7.46), $\log(gf)$ ($-$4.65), microturbulence (0.4 \kms), and
$V\cdot\sin(i)$ as in Table~\ref{tab:weaklins}. But $V_{\rm mac}$ was
adjusted to obtain optimum fits.  For Profiles 1-4, the best-fitting values
of $V_{\rm mac}$ were 1.1, 2.7, 3.9,  and 3.0 \kms.

\subsection{Moderately weak lines\label{sec:modwk}}

A similar experiment was performed for slightly larger equivalent widths.
We froze the abundance (7.46), $V_{\rm mac}$ (2.4 \kms), and  $V\cdot\sin(i)$
(0.0 \kms), and calculated equivalent widths and profiles
for $\log(gf) = -4.30$, $-$4.10,
and $-$3.96.  The corresponding equivalent widths were 10.5, 15.5, and 20.3 m\AA.
We then tried to match these profiles by varying the abundances and $V_{\rm mac}$.
Relevant parameters for the fitted profiles
are shown in Table~\ref{tab:modwk}.   The corresponding
figures are not shown as they closely resemble the fits of Fig.~\ref{fig:All4}.

As we can see from Table~\ref{tab:modwk}, the accuracy of the abundance 
(determined by the difference $ A_{\rm original}-A_{\rm fit}$) decreases
with the increase of the equivalent width of the moderately weak lines. The
uncertainties of the microturbulence can cause an error about 0.02--0.05~dex
in  the abundance derived from the profile fit of weak line with equivalent
width of 10--20~m\AA.  This is well within the typical accuracy of the
oscillator strengths.
 \begin{table}
 \caption{Parameters of the best weak ($<$21 m\AA) profile fits
 discussed in Section~\ref{sec:modwk}.  Equivalent
 widths of the fitted profiles ($W_{\rm fit}$) are the same (to 0.1 m\AA) as
 those of the original calculation (with $A_{\rm original} = 7.46$) apart from Profile 3.
 }
   \label{tab:modwk}
   \begin{tabular}{c c c c c c c} \hline
   Pro. & $W_{\rm fit}$&$\log (gf)$& $\xi_t$& $A_{\rm fit}$ &$V_{\rm mac}$&  $\chi^2/n$   \\
        & m\AA\, &           & \kms   &     & \kms        &               \\
        \hline
     1  & 5.0    &-4.65      &  0.4   &  7.46   &  2.7  &  $1.7\cdot 10^{-7}$          \\
     2  &10.5    &-4.30      &  0.4   &  7.48   &  2.8  &  $3.4\cdot 10^{-7}$        \\
     3  &15.7    &-4.10      &  0.4   &  7.50   &  2.8  &  $1.7\cdot 10^{-7}$          \\
     4  &20.3    &-3.96      &  0.4   &  7.51   &  2.8  &  $4.0\cdot 10^{-7}$         \\ \hline
  \end{tabular}
  \end{table}

  \section{Intermediate-strength lines}
  \label{sec:intstr}

  \begin{figure}
    \centering
  \includegraphics[scale=0.75]{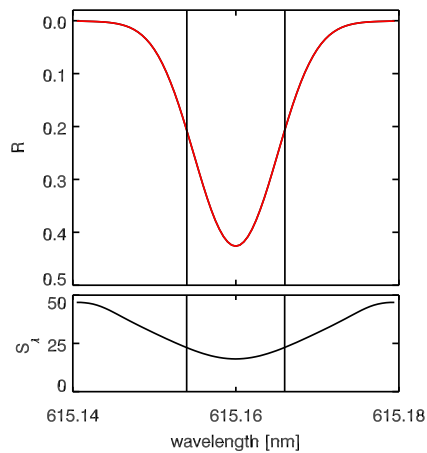}
   \caption{Sensitivity  (black) and line depth profile (red) of
   calculated  53.4 m\AA\, line in the solar flux.  Note that the
   most sensitive parts of the profile to abundance lie where
   the line is very weak, or in a steep part of the profile.
   Vertical lines indicate the sensitivity level of 50\% of the maximum
    $S_\lambda$ shown.}
  \label {fig:modS}
  \end{figure}
  \begin{figure}
    \centering
   \includegraphics[scale=0.95]{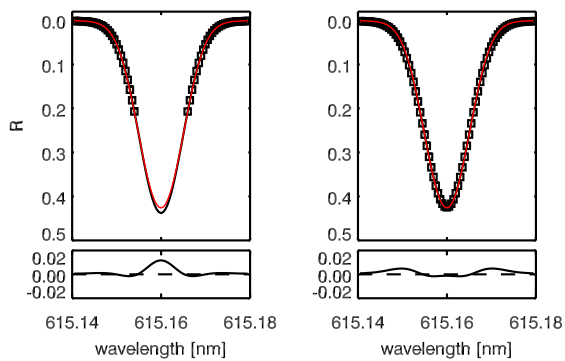}
\caption{A  best fit to the original (red) profile may be made for sensitive
points (left, squares) with $\xi_t = 0.4$, $V_{\rm mac}=2.7$~\kms, $A =
7.74$, $\chi^2 /n = 8.8\cdot10^{-7}$ as well as for whole profile (right,
with $\xi_t = 0.4$, $V_{\rm mac}=2.8$~\kms, $A=7.72$, $\chi^2 /n =
4.8\cdot10^{-6}$ .
     }
  \label {fig:modP}
  \end{figure}


It is our opinion that abundances should be based on weak lines ($W_\lambda
\le {\approx } 20$ m\AA), whenever reliable oscillator strengths are
available for them. In this and the next section we turn to a consideration
of intermediate-strength, and strong lines, and illustrate their
shortcomings.  However, it would be a mistake to conclude that there is no
abundance information in such features.  In spite of their shortcomings it
is sometimes necessary to work with them.

Curves of growth tell us that intermediate-strength lines are more sensitive
to broadening mechanisms than abundance.  For such features, a knowledge of
sensitive points is not a great advantage. Apart from the risky regions with
very small line depth, the more sensitive points lie in the steepest parts
of the line profiles, where accurate measurements are difficult. Numerical
experiments show that an underestimated microturbulence can be compensated
by adjustments to the abundance and $V_{\rm mac}$ in such a way as to
reproduce the original profile to an entirely acceptable accuracy.  Yet the
abundance difference could be more than 0.2~dex different from the correct
value.  This is illustrated in Figs.~\ref{fig:modS}, and ~\ref{fig:modP}.

Figure~\ref{fig:modS} shows the sensitivity function and line-depth profile
for a 53.4 m\AA\, line calculated in the solar flux with  MARCS model. The
following parameters were used in the calculation:  $A=7.46$, $\log(gf) =
-3.29$, $\xi_t=1.4$~\kms, $V \sin i=0$. The profile was convolved with
$V_{\rm mac}=2.4$~\kms.

In the  numerical experiment, we set $\xi_t$ to 0.4~\kms, and attempted to
reproduce the original profile by adjusting the abundance and $V_{\rm mac}$.
Fig.~\ref{fig:modP} shows the best fit for the sensitive points is excellent
($A=7.74$, $\chi^2 /n =8.8 \cdot 10^{-7}$), while for whole profile is fair
($A=7.72$, $\chi^2 /n =4.8 \cdot 10^{-6}$), but should satisfy an abundance
worker with a realistic sense of the uncertainties of this work.  Note that
the `new' abundance is greater by 0.26--0.28 dex than the value originally
assumed.

  \section{Strong lines}
  \label{sec:strong}

\begin{figure}
    \centering
     \includegraphics[scale=0.8]{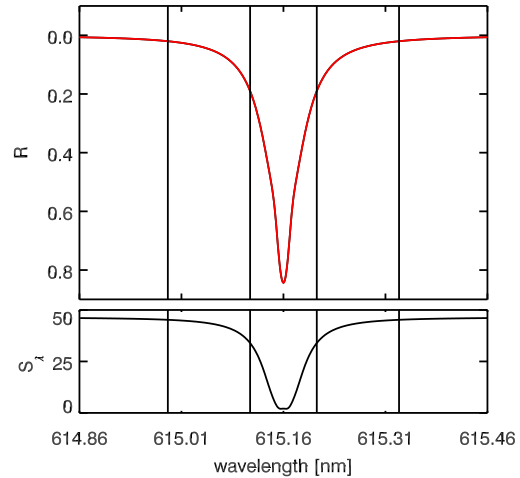}
\caption{Sensitivity function (black) and depth profile (red) of a 642.3
m\AA\, line.  The strips on either side of the line center show  regions of
the sensitive points, that was chosen with the sensitivity is greater than
75\% of its  maximum value (Table~\ref{tab:Parms7}, Pro.~1).
     }
  \label {fig:strongS}
  \end{figure}

\begin{figure}
    \centering
        \includegraphics[scale=1]{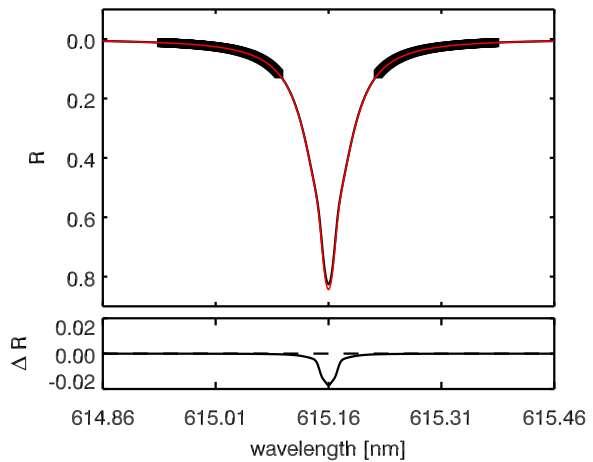}
   \caption{A  best fit to the original (red) profile of a strong line
may be made for sensitive points (black squares) with $E_6 =1.75 $,
$A=7.24$, $\chi^2 /n = 1.8\cdot10^{-8}$ (Table~\ref{tab:Parms7}, Pro.~2).
     }
  \label {fig:strongP}
  \end{figure}

The wings of strong lines are potentially useful for abundances subject to
the severe qualification that the broadening mechanisms be accurately known.
The wing strength of a line depends directly on the product of the
abundance, the oscillator strength, and the damping constant.  Thus, the
damping constant must be as accurately known as the oscillator strength.  We
illustrate this in Figs.~\ref{fig:strongS}, and ~\ref{fig:strongP}, which
are similar to the figures of Section~\ref{sec:intstr}.

We used the same Fe \I \, line as  Section~\ref{sec:intstr}, but changed the
$\log(gf)$ from $-$3.29 to 0.00 and the wavelength step from 5 to 10~m\AA.
With $A=7.46$, $\xi_t = 1.4$~\kms, $V_{\rm mac}=2.4$~\kms, we obtain the
642.3~m\AA\, line shown in Fig.~\ref{fig:strongS}.  For this line 75\% of
the maximum sensitivity shown corresponds $R=0.18$.  We can use the
sensitive points in the wing regions of $0.02<R<0.18$ that are  in a more
gradually-slopping part of the profile than was the case for
Fig.~\ref{fig:modP}.

  \begin{table}
  \caption{Parameters for Figs.~\ref{fig:strongS}, ~\ref{fig:strongP}, and ~\ref{fig:Vt34}.
  Log($gf$) and $V\cdot\sin(i)$  are 0.00 for all three profiles.
  $E_6$ is an enhancement factor  to the collisional damping.
  }
  \label{tab:Parms7}
   \begin{tabular}{c c c c c c c} \hline
   Pro. & $W$    & $\xi_t$& $A$   &$V_{\rm mac}$&  $E_6$  &  $\chi^2/n$  \\
        & m\AA\, & \kms   &       & \kms        &             &\\
        \hline
     1  & 642.3  &  1.4   &  7.46 &  2.4        & 1.00     &\\
     2  & 636.5  &  1.4   &  7.24 &  2.4        & 1.75 & $1.8\cdot 10^{-8}$ \\
     3  & 692.4  &  3.4   &  7.46 &  2.4        &1.00     &$3.2\cdot 10^{-7}$ \\
 \hline
  \end{tabular}
  \end{table}

To demonstrate the degeneracy of abundance and damping,  we set the  damping
constant ($C_6$) increased by a factor $E_6=1.75$ and  tried to compensate
that by changing abundance.  To get the best fit for this new line profile,
the abundance was decreased by 0.22 dex (Fig.~\ref{fig:strongP}).
Good abundances {\em can} be determined from the wings of
strong lines when the damping constants are accurately known.
If rotation and macroturbulence are significance, these factors
must also be known.

In differential abundance work the often uncertain damping constants and
oscillator strengths may largely cancel.  The degree to which these factors
could be important depends on how closely the differenced stars resemble one
another.  Residual effects of rotation or macroturbulence must be carefully
considered.

\begin{figure}
    \centering
        \includegraphics[scale=1]{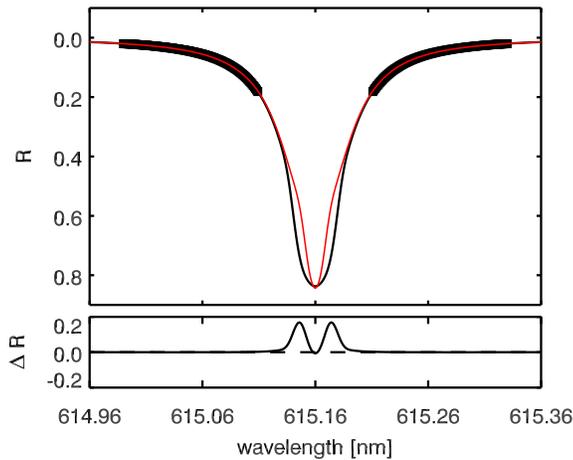}
   \caption{This strong line calculation (black) uses all parameters
   including the abundance and damping the same as the (red, Table~\ref{tab:Parms7}, Pro.~1)
   calculation, but the microturbulence has been changed from 1.4
   to 3.4 \kms (Table~\ref{tab:Parms7}, Pro.~3).
     }
  \label {fig:Vt34}
  \end{figure}

We now show (Fig.~\ref{fig:Vt34}) the case of a strong line where the core
is not well fit, but an excellent  abundance  would be obtained from a fit
to the sensitive points.
The sensitive region contains 313 points. The  best fit to the sensitive
points gives the original $A=7.46$ and a small $\chi^2/n$ of 3.2$\cdot
10^{-7}$. If the central region of the profile is included, there are  799
points, and the fit is obviously less good; $\chi^2/n = 7.2\cdot 10^{-4}$.

In this section we have demonstrated that a fit to a strong
line--including sensitive points--does not necessarily mean that a correct
abundance will be obtained.  On the other hand we show that even if the core
of a strong line is poorly fit, an accurate abundance can be obtained
when the broadening is accurately known.

\section{Summary\label{sec:summary}}

Not all portions of a stellar spectrum are equally sensitive to abundances.
In this paper we showed that a good abundance not only depends on the choice
of accurate atomic parameters, but the selection of features that are
sensitive to abundances. We proposed a scheme to find abundance-sensitive
regions.  By far, the most useful features for abundances are weak lines
with reliable $gf$-values. The use of weak features requires accurate
continuum adjustment, which is rarely as accurate as 0.5 per cent.
Unidentified blends or  blends with inaccurate atomic parameters in the line
wings do not allow the use of most profiles in the abundance analysis.
However, we can select sensitive points in the profile parts are free from
the blends or omit some points near the continuum  where these effects are
significant.

We show that moderately strong lines, in spite of having sensitive regions,
are usually less good for abundance work due to their high sensitivity to
turbulence.  Today, the application of 1D atmosphere models and classical
concepts of micro and macroturbulence  are widely used and the problem of
uncertainty of the synthetic profiles due to the micro and macroturbulence
remains for sensitive profile points. While the line profile results
obtained with 3D hydrodynamical (HD) stellar atmosphere models without any
need for micro or macroturbulence are not without problems. Theoretical profiles
of the moderate lines computed in 3D HD systematically underestimate the
line width and that  some additional work on improving the atmospheric
velocity field is still required
 \citep{2014arXiv1405.0287S}.

Strong lines can be useful, especially in differential abundance work. The
wings of very strong lines are insensitive to turbulence parameters and the
NLTE-effects. They can be used for profile fits when the damping constants
are well known.

We propose to using the sensitive points  but recognize that other regions
of the spectrum still contain abundance information.

\subsection*{Acknowledgements}

We thank an unknown referee for patience, and suggestions to improve the
paper.

 

\end{document}